\documentclass[a4paper,11pt]{JHEP3}
\usepackage{fancyheadings}
\usepackage[dvips]{graphicx}
\usepackage{ifthen}
\usepackage{amsmath}
\usepackage{epsfig}
\usepackage[errorshow]{tracefnt}
\newcommand\nn{\nonumber}

\newcommand\beal{\begin{align}}
\newcommand\eeal{\end{align}}
\newcommand\benu{\begin{enumerate}}
\newcommand\eenu{\end{enumerate}}
\newcommand\bit{\begin{itemize}}
\newcommand\eit{\end{itemize}}

\newcommand{\be}{\begin{equation}}
\newcommand{\la}{\label}
\newcommand{\ee}{\end{equation}}
\newcommand{\bd}{\begin{displaymath}}
\newcommand{\ed}{\end{displaymath}}

\newcommand\cw{{\cal W}}

\newcommand\g{\gamma}
\newcommand\de{\delta}

\newcommand\e{\epsilon}

\newcommand\G{\Gamma}
\newcommand\Om{\Omega}

\title{Supersymmetric $AdS_4$ compactifications of IIA supergravity}

\author{Dieter L\"{u}st \\
Max-Planck-Institut f\"{u}r Physik --Theorie\\
F\"{o}hringer Ring 6, 80805 M\"{u}nchen, Germany\\ \&\\
Department f\"{u}r Physik, Ludwig-Maximillians-Universit\"{a}t\\
Theresienstr. 37, 80333 M\"{u}nchen, Germany 
\\E-mail: \email{luest@mppmu.mpg.de, luest@theorie.physik.uni-muenchen.de}
}
\author{Dimitrios Tsimpis \\
Max-Planck-Institut f\"{u}r Physik --Theorie\\
F\"{o}hringer Ring 6,  80805 M\"{u}nchen, Germany\\
E-mail: \email{tsimpis@mppmu.mpg.de}
}

\abstract{We derive necessary and sufficient conditions 
for ${\cal N}=1$ compactifications of 
(massive) IIA supergravity to $AdS_4$ in the 
language of $SU(3)$ structures. We find new solutions 
characterized by constant dilaton and nonzero fluxes for all 
form fields. All fluxes are given 
in terms of the geometrical data of the internal compact space.   
The latter is constrained to belong 
to a special class of half-flat manifolds. }

\preprint{MPP-2004-173}
\begin{document}

\section{Introduction and Summary}

Flux compactifications \cite{1,2,3,4,4.1,5,5.1,6,7,8,8.1,8.2,9,10} 
are currently being studied intensively, not least for their 
potential phenomenological interest. Perhaps their most attractive feature is that 
they allow the possibility of fixing (part of) the geometrical moduli of the internal space. 
Moreover fluxes generally source warp factors, which can provide a mechanism 
 for generating hierarchies.

In many cases it suffices to work in an approximation where the back-reaction of the 
fluxes on the internal manifold is ignored. In ${\cal N}=2$ 
type II compactifications to four-dimensional 
Minkowski space, for example, one continues to treat the internal manifold as 
if it were a Calabi-Yau, 
even after giving expectation values to the antisymmetric tensors along the internal directions. 
This situation is usually described as `Calabi-Yau with fluxes' despite the fact that it 
does not correspond to a true supergravity solution. 

This approach is motivated partly by the fact that the physics community has 
grown particularly fond of 
Calabi-Yau manifolds, on which one can use familiar tools from algebraic geometry. 
Once fluxes are turned on, however, the internal manifold is deformed away from the 
Calabi-Yau point and, generically, it will even cease to be complex. The manifolds which appear 
naturally in the setup of the present paper, for example, belong to the class of 
{\it half-flat} manifolds \cite{cs} also known as {\it half-integrable}   
\cite{cswann}  
about which little is known. Notably, they appear in the mirror-symmetric 
picture of `Calabi-Yau with fluxes' compactifications \cite{121,122}. 

For many practical purposes it has proven fruitful to work within the approximation described 
above and to ignore the back-reaction of the fluxes. Nevertheless it would still 
be desirable to obtain exact results, corresponding to 
genuine supergravity solutions. Ideally, one would like to be able 
to classify and systematically construct 
concrete examples of internal manifolds satisfying the requirements for 
a consistent (supersymmetric) vacuum with fluxes. This task,  however, is 
well beyond our current technology. 
The knowledge of exact solutions may provide clues as to how to set up 
some kind of perturbative expansion for which the order parameter would be the flux.  
`Calabi-Yau with fluxes' would then correspond to the leading-order term in this expansion.

The subject of supersymmetric 
supergravity compactifications is of course not a new one. 
More recently, it has become clear that the most suitable language 
for the description of flux compactifications is that of $G$-structures. 
There is already a considerable amount of literature on the subject, see 
\cite{11,12,letal,13,14,15,16,17,18,19,dp,20,21,22,23,24} 
for an inexhaustive list of references. 

The necessary conditions for ${\cal N}=1$ 
IIA (and IIB) supergravity compactifications to four-dimensional Minkowski space were given 
in the language of $SU(3)$ structures in \cite{getala, getalb}. The authors of these 
references examined the conditions for unbroken supersymmetry 
but did not impose the Bianchi identities or the equations of motion. 
The latter were taken into account in reference \cite{dp}. 
A class of ${\cal N}=1$ compactifications of Romans' supergravity to 
$AdS_4$ was given in \cite{bca, bc}, corresponding to the internal manifold being 
{\it nearly-K\"{a}hler}. In the massless limit these solutions reduce to 
four-dimensional Minkowski space times a Calabi-Yau, and all fluxes vanish. 

In the present paper we 
derive the necessary and sufficient conditions for ${\cal N}=1$ compactifications 
of (massive) IIA supergravity to $AdS_4$ in the context of $SU(3)$ structures, and 
we find a new class of solutions.  
These are characterized by 
constant dilaton and scalar fluxes (the term `scalar' 
referring to the decomposition 
in terms of the $SU(3)$ structure group) 
turned on for all form fields. In addition, the (massive) two-form 
can have a nonzero primitive piece. 
The intrinsic torsion of the internal manifold $X_6$ is constrained 
by supersymmetry and the Bianchi identities to lie in 
\beal
\tau\in\cw^-_1\oplus\cw^-_2
\label{i1}
\end{align}
and therefore $X_6$ is a half-flat manifold.  
Recall that the latter is a manifold whose intrinsic torsion is contained in 
$\cw_1^-\oplus\cw_2^-\oplus\cw_3$. 
Within the class of half-flat manifolds, our solutions 
are `orthogonal' to the example based on the 
Iwasawa manifold encountered in \cite{letal}, in the sense 
that for the latter the intrinsic torsion is entirely contained in $\cw_3$. 
In addition to (\ref{i1}), 
the Bianchi identities require the exterior derivative of 
$\cw_2^-$ to be proportional to the real part of the 
$(3,0)$ form on the manifold $X_6$, 
\beal
d\cw_2^-\propto Re(\Omega)~.
\label{i2}
\end{align}
Moreover, all form field components are expressible in terms of the geometrical  
data of $X_6$. The equations-of-motion are then 
satisfied with no further requirements. 

In contrast to \cite{bca, bc} our solutions reduce in the massless limit 
to $AdS_4$ times a six-dimensional manifod $X_6$ of the type 
(\ref{i1}, \ref{i2}). The lift to M-theory can then 
be taken, and leads to a seven-dimensional 
internal manifold which is a twisted circle fibration 
over $X_6$. This theory is expected to admit a three-dimensional conformal 
field theory dual, see \cite{js} for a recent discussion and \cite{nunez} 
for earlier work on this subject.

In section \ref{examples} we examine examples of six-dimensional 
manifolds with properties (\ref{i1},\ref{i2}). 
More specifically, in section \ref{t2overk3} we construct a  class of examples 
of six-dimensional manifolds which are 
$T^2$ fibrations over K3. In addition, in section \ref{iwasawa} 
we examine a one-parameter family of 
examples based on the Iwasawa manifold. 
To our knowledge, of all the 
nilmanifolds considered in the mathematical literature 
\cite{ags, cs, cswann, cf}, 
this is the only one  for which (\ref{i1}) holds. This 
example turns 
out to be a degeneration of the case considered in section \ref{t2overk3}, 
whereby the $K3$ base is replaced by a $T^4$.  
Although the examples of section \ref{examples} do satisfy 
equations (\ref{i1},\ref{i2}), they fail to reproduce the specific 
constant of proportionality between $d\cw_2^-$ and the real part of the $(3,0)$ form 
required by the Bianchi identities. 
The construction of more examples of the type (\ref{i1},\ref{i2}) will 
have to await further developments in the mathematical literature.

In section \ref{kiki} we review Romans' ten-dimensional supergravity and examine 
the integrability of the supersymmetry variations. We conclude that 
imposing supersymmetry, the Bianchi identities and the form equations of motion 
suffices for the dilaton and Einstein equations to be automatically satisfied. 
Similar statements have previously appeared in the literature 
(see \cite{13},\cite{bandos} for a discussion 
in the context of eleven-dimensional supergravity) but, 
to our knowledge, not in the 
context of the present paper. 
In section \ref{reduction} we 
reduce on $AdS_4\times X_6$ 
taking into account the fact that the internal manifold possesses an $SU(3)$ structure. 
The necessary and sufficient conditions for ${\cal N}=1$ supersymmetry 
are derived in section \ref{lisp}, where our solutions are presented. 
Section \ref{examples} is devoted to the explicit construction of examples. 
The two appendices contain our conventions and some useful technical results.

\section{Massive IIA}
\label{kiki}

This section contains a review of Romans massive supergravity \cite{roma}. 
It is included here in order to establish notation and conventions. 

The bosonic part of the action reads
\beal
{\cal L}
=\int \{ R{}* 1-\frac{1}{2}d\phi\wedge*d\phi&-\frac{1}{2}e^{\phi/2}
G\wedge *G -\frac{1}{2}e^{-\phi}H\wedge * H 
-2m^2 e^{3\phi/2}B'\wedge *B'\nn\\
&+\frac{1}{2}(dC')^{2}\wedge B'+\frac{m}{3}dC'\wedge(B')^{3}
+\frac{m^2}{10}(B')^{5}-2m^2 e^{5\phi/2}*1\} ~,
\la{lagr}
\end{align}
where
\be
H=dB'
\label{prwtn}
\end{equation}
and
\be
G=dC'+m(B')^2~.
\end{equation}
These forms obey the Bianchi identities 
\beal
dH&=0\nn\\
dG&=2mB'\wedge H~.
\label{bianchi1}
\end{align}
Note that we are using `superspace' conventions for the forms:
\beal
A_{(n)}&=\frac{1}{n!}dx^{M_n}\wedge\dots dx^{M_1}A_{M_1\dots M_n}\nn\\
d(A_{(n)}\wedge B_{(q)})&=A_{(n)}\wedge dB_{(q)}+(-)^q dA_{(n)}\wedge B_{(q)}~.
\end{align}
Recall that there is no (known) covariant lift of massive IIA to eleven dimensions. 
To make contact with the massless IIA supergravity of \cite{gp, cw, hn},  
one introduces a St\"{u}ckelberg gauge  potential $A$ so that
\beal
m B'&=m B+\frac{1}{2}F\nn\\
m C'&=m C-\frac{1}{4}A\wedge F\nn\\
F=dA; ~~~~~~ H=dB&; ~~~~~~G=dC+B\wedge F+mB^2~.
\label{masslesslimit}
\end{align}
The Bianchi identities of the forms read
\beal
dF&=0\nn\\
dH&=0\nn\\
dG&=H\wedge F+2m B\wedge H~.
\label{bianchi2}
\end{align}
After introducing the St\"{u}ckelberg field, the theory is invariant under
\beal
A&\rightarrow A+m\Lambda\nn\\
B&\rightarrow B-\frac{1}{2}d\Lambda\nn\\
C&\rightarrow C+\frac{1}{2}A\wedge d\Lambda+\frac{m}{4}
\Lambda\wedge d\Lambda~.
\end{align}
Moreover, one can check that in terms of the 
fields $A,B,C$ the Chern-Simons terms in (\ref{lagr}) can 
be rewritten, up to a total derivative, as
\be
CS=\frac{1}{2}dC^2\wedge B+\frac{1}{2}dC\wedge B^2\wedge F+
\frac{1}{6}B^3\wedge F^2
+\frac{m}{3}dC\wedge B^3+\frac{m}{4}B^4\wedge F+\frac{m^2}{10}B^5~,
\nn
\end{equation}
so that the $m\rightarrow 0$ limit can now be taken and the 
Lagrangian reduces to the one of massless IIA supergravity.

\subsection*{Equations of motion}

The equations of motion that follow from (\ref{lagr}) are
\beal
0&=R_{MN}-\frac{1}{2}\nabla_M\phi\nabla_N\phi-\frac{1}{12}e^{\phi/2}
G_{MPQR}G_{N}{}^{PQR}+\frac{1}{128}e^{\phi/2}g_{MN}G^2\nn\\
&~~~~~-\frac{1}{4}e^{-\phi}
H_{MPQ}H_{N}{}^{PQ}
+\frac{1}{48}e^{-\phi}g_{MN}H^2\nn\\
&~~~~~-2m^2e^{3\phi/2}
B'_{MP}B'_{N}{}^{P}+\frac{m^2}{8}e^{3\phi/2}g_{MN}(B')^2
-\frac{m^2}{4}e^{5\phi/2}g_{MN}\\
0&=\nabla^2\phi-\frac{1}{96}e^{\phi/2}G^2+\frac{1}{12}e^{-\phi}H^2
-\frac{3m^2}{2}e^{3\phi/2}(B')^2-5m^2e^{5\phi/2}\\
\la{heq}
0&=d(e^{-\phi}*H)-\frac{1}{2}G\wedge G+2m~e^{\phi/2}B'\wedge*G
+4m^2e^{3\phi/2} *B'\\
\la{geq}
0&=d(e^{\phi/2}{}*G)-H\wedge G~.
\end{align}
Note that the integrability condition following from 
(\ref{heq}),
\be
0=e^{\phi/2}H\wedge *G+2m~d(e^{3\phi/2}*B')~, 
\end{equation}
becomes in the massless limit the equation of motion for the gauge field.

\subsection*{Supersymmetry}

The gravitino and dilatino supersymmetry variations read
\beal
\delta\Psi_M={\cal D}_M\e
\la{gravitino}
\end{align}
and
\beal
\de\lambda=\{
-\frac{1}{2}\G^M\nabla_M\phi&-\frac{5m~e^{5\phi/4}}{4}
+\frac{3m~e^{3\phi/4}}{8}B'_{MN}\G^{MN}\G_{11}\nn\\
&+\frac{e^{-\phi/2}}{24}H_{MNP}\G^{MNP}\G_{11}
-\frac{e^{\phi/4}}{192}G_{MNPQ}\G^{MNPQ}
\}\e~,
\la{dilatino}
\end{align}
where the supercovariant derivative is given by 
\beal
{\cal D}_M&:=\nabla_M-\frac{m~e^{5\phi/4}}{16}\Gamma_M
-\frac{m~e^{3\phi/4}}{32}
B'_{NP}(\G_M{}^{NP}-14\de_M{}^N\G^P)\G_{11}\nn\\
&+\frac{e^{-\phi/2}}{96}H_{NPQ}(\G_M{}^{NPQ}-9\de_M{}^N\G^{PQ})\G_{11}
+\frac{e^{\phi/4}}{256}G_{NPQR}(\G_M{}^{NPQR}
-\frac{20}{3}\de_M{}^N\G^{PQR})
\end{align}
and $\epsilon$ is the susy parameter. One can transform to the string frame 
by rescaling $e_A{}^M\rightarrow e^{\phi/4}e_A{}^M$.

\subsection*{Integrability}

We will now argue that in a 
purely bosonic supersymmetric background 
the vanishing of the supersymmetric variations of the fermions 
together with the Bianchi identities and equations-of-motion for the forms 
imply (under a further mild assumption 
which is satisfied by the compactifications considered in 
the present paper --see below) 
the dilaton and Einstein equations. To our knowledge, this is the 
first time this has been shown in the context of the present paper. 
For the purposes of this subsection we set $B'=0$ for simplicity 
of presentation. The conclusion 
does not change by introducing a nonzero $B'$-field. We have  
found \cite{gran} extremely useful in the following computation. 

In a bosonic supersymmetric background, the supersymmetric variations 
of the fermions have to vanish. Assuming this to be the case, 
one can use the gravitino variation to obtain an expression 
for the commutator of two supercovariant derivatives acting on the susy 
parameter,
\be
2{\cal D}_{[M}{\cal D}_{N]}\e=\{\frac{1}{4}R_{MNPQ}\G^{PQ}+\dots\}\e=0~.
\la{0}
\end{equation}
Furthermore, the vanishing of the dilatino variation gives
\be
\frac{5m}{4}e^{5\phi/4}\e=\{-\frac{1}{2}\G^M\nabla_M\phi-
\frac{e^{\phi/4}}{192}G_{MNPQ}\G^{MNPQ}\}\e~.
\la{1}
\end{equation}
Taking the `square' of the above expression we get,
\be
\frac{1}{4}(\nabla\phi)^2\e= \{
\frac{25m^2}{16}e^{5\phi/2}+\dots\}\e~.
\la{2}
\end{equation}
Multiplying (\ref{0}) by $\G^N$ and substituting (\ref{1},\ref{2}) we obtain 
\beal
0=\{&-\frac{1}{2}\G^N(R_{MN}-\frac{1}{2}\nabla_M\phi\nabla_N\phi
-\frac{1}{12}e^{\phi/2}
G_{MPQR}G_{N}{}^{PQR}+\frac{1}{128}e^{\phi/2}G^2g_{MN}-\frac{m^2}{4}
e^{5\phi/2}g_{MN} )\nn\\
&+\frac{e^{\phi/4}}{256}\G_M{}^{IJKLP}\nabla_{[I}G_{JKLP]}
-\frac{25e^{\phi/4}}{768}\G^{IJKL}\nabla_{[M}G_{IJKL]}\nn\\
&+\frac{e^{-\phi/4}}{64}\G_M{}^{IJK}\nabla^L(e^{\phi/2} G_{LIJK})
-\frac{5e^{-\phi/4}}{64}\G^{IJ}\nabla^L(e^{\phi/2}G_{LMIJ})\nn\\
&+\frac{e^{\phi/2}}{4608}(
\G_{MI_1\dots I_8}G^{I_1\dots I_4}G^{I_5\dots I_8}
-\frac{1}{4}\G_M \G_{I_1\dots I_8}G^{I_1\dots I_4}G^{I_5\dots I_8}
)~\}\e~.
\la{JSB}
\end{align}
Imposing the Bianchi identities  
and equations of motion for the forms, 
equation (\ref{JSB}) takes the form 
$E_{MN}\G^N\e =0$, 
where $E_{MN}=0$ are the Einstein equations. Multiplying 
this by $E_{MP}\G^{P}$ implies $E_{MN}E_M{}^{N}=0$ 
(no summation over $M$). 
Furthermore, if $E_{M0}$ vanishes for $M\neq 0$, 
the remaining terms 
in the sum are positive-definite 
and one obtains $E_{MN}=0$, $M\neq 0$. Finally, $E_{0N}E_0{}^{N}=0$ 
implies $E_{00}=0$ and therefore $E_{MN}=0$, for all $M,N$. 
A similar statement can be made for the dilaton equation, 
as one can see by acting on the dilatino variation with $\Gamma^N\nabla_N$.

In conclusion, supersymmetry together 
with the Bianchi identities and equations of motion for the forms 
imply that 
the dilaton and Einstein equations are satisfied, provided $E_{0M}=0$ for $M\neq 0$.

\section{$M_{1,3}\times X_6$ backgrounds}
\label{reduction}

\subsection{Supersymmetry}

Let us now assume that spacetime
is of the form of a warped product 
$M_{1,3}\times_{\omega} X_6$, where $M_{1,3}$ is 
$AdS_4$ (or $\mathbb{R}^{1,3}$) and $X_6$ is a compact manifold. 
The ten dimensional metric reads
\be
g_{MN}(x,y) = 
\left(
\begin{array}{cc}
\Delta^2(y) \hat{g}_{\mu\nu}(x) & 0 \\ 
0 & g_{mn}(y)
\end{array}
\right)
~,
\end{equation}
where $x$ is a coordinate on $M_{1,3}$ and $y$ is a coordinate on $X_6$. 
We will also assume that the forms have nonzero 
$y$-dependent components 
along the internal directions, except for the four-form which will be 
allowed to have an additional component proportional to the volume 
of $M_{1,3}$
\be
G_{\mu\nu\kappa\lambda}=\sqrt{g_{4}}f(y)\varepsilon_{\mu\nu\kappa\lambda}~,
\end{equation}
where $f$ is a scalar on $X_6$.
Note that with these assumptions 
the $E_{M0}=0$ for $M\neq 0$ condition is satisfied, and therefore 
we need only check supersymmetry 
the Bianchi identities and the equations of motion for the forms.

The requirement of $\mathcal{N}=1$  
supersymmetry in $4d$ (4 real supercharges) implies the existence of
 a globally defined 
complex spinor $\eta$ on $X_6$. 
As a consequence the structure group of $X_6$ reduces to $SU(3)$, 
as explained in the following subsection in more detail. In addition, on
 $M_{1,3}$ there is a pair of Weyl spinors (related by complex conjugation), 
each of which 
satisfies the Killing equation
\be
\hat{\nabla}_\mu\theta_+=W\hat{\g}_\mu\theta_-~; 
~~~~~~~
\hat{\nabla}_\mu\theta_-= W^*\hat{\g}_\mu\theta_+ ~,
\la{ads}
\end{equation}
where hatted quantities are computed using   
the metric $\hat{g}_{\mu\nu}$, and 
the complex constant $W$ is related to the scalar curvature $\hat{R}$ 
of $M_{1,3}$ through $\hat{R}=-24|W|^2$. Spinor conventions are given in appendix A. 

The ten-dimensional spinor $\e=\e_++\e_-$ decomposes as 
\be
\e=(\alpha\theta_+\otimes\eta_+-\alpha^*\theta_-\otimes\eta_-)+(
\beta \theta_+\otimes\eta_--\beta^*\theta_-\otimes\eta_+)~,
\la{spinoransatz}
\end{equation}
where $\alpha$, $\beta$ are complex functions on $X_6$, undetermined 
at this stage. The spinor $\e$ thus defined 
is Majorana.

Substituting these Ans\"{a}tze in the supersymmetry 
transformations we obtain
\beal
0&=\alpha\nabla_m\eta_+
+\partial_m\alpha\eta_+
+\alpha\frac{e^{-\phi/2}}{96}H_{npq}
(\g_m{}^{npq}-9\delta_m{}^n\g^{pq})\eta_+
-\beta\frac{me^{5\phi/4}}{16}\g_m\eta_-
+3i\beta f\frac{e^{\phi/4}}{32}\g_m\eta_-
\nn\\
&+\beta\frac{me^{3\phi/4}}{32}B'_{np}
(\g_m{}^{np}-14\delta_m{}^n\g^p)\eta_-
+\beta\frac{e^{\phi/4}}{256}G_{npqr}(\g_m{}^{npqr}-\frac{20}{3}\delta_m{}^n
\g^{pqr})\eta_-
\la{eva}\\
0&=\beta^*\nabla_m\eta_+
+\partial_m\beta^*\eta_+
-\beta^*\frac{e^{-\phi/2}}{96}H_{npq}
(\g_m{}^{npq}-9\delta_m{}^n\g^{pq})\eta_+
+\alpha^*\frac{me^{5\phi/4}}{16}\g_m\eta_-
+3i\alpha^* f\frac{e^{\phi/4}}{32}\g_m\eta_-
\nn\\
&+\alpha^*\frac{me^{3\phi/4}}{32}B'_{np}
(\g_m{}^{np}-14\delta_m{}^n\g^p)\eta_-
-\alpha^*\frac{e^{\phi/4}}{256}G_{npqr}(\g_m{}^{npqr}-\frac{20}{3}\delta_m{}^n
\g^{pqr})\eta_-~,
\la{duo}
\end{align}
from the `internal' components of the gravitino variation and
\beal
0&=\alpha\Delta^{-1}W\eta_++\beta^*\frac{me^{5\phi/4}}{16}\eta_+
-5i\beta^*f\frac{e^{\phi/4}}{32}\eta_+
-\beta^*\frac{me^{3\phi/4}}{32}B'_{mn}\g^{mn}\eta_+\nn\\
&+\alpha^*\frac{e^{-\phi/2}}{96}H_{mnp}\g^{mnp}\eta_-
-\beta^*\frac{e^{\phi/4}}{256}G_{mnpq}\g^{mnpq}\eta_+
-\frac{1}{2}\alpha^*\partial_m(ln\Delta)\g^m\eta_-
\la{tria}\\
0&=\beta^*\Delta^{-1}W^*\eta_++\alpha\frac{me^{5\phi/4}}{16}\eta_+
+5i\alpha f\frac{e^{\phi/4}}{32}\eta_+
+\alpha\frac{me^{3\phi/4}}{32}B'_{mn}\g^{mn}\eta_+\nn\\
&+\beta\frac{e^{-\phi/2}}{96}H_{mnp}\g^{mnp}\eta_-
-\alpha\frac{e^{\phi/4}}{256}G_{mnpq}\g^{mnpq}\eta_+
+\frac{1}{2}\beta\partial_m(ln\Delta)\g^m\eta_-
~,
\la{tessera}
\end{align}
from the noncompact piece. Note that these equations are complex. 
Similarly from the dilatino we obtain
\beal
0&=\frac{1}{2}\alpha^*\partial_m\phi \g^m  \eta_- -
\alpha^*\frac{e^{-\phi/2}}{24}H_{mnp}\g^{mnp}\eta_-
-\beta^*\frac{5me^{5\phi/4}}{4}\eta_+\nn\\
& +i\beta^* f\frac{e^{\phi/4}}{8}\eta_+
-\beta^*\frac{3me^{3\phi/4}}{8}B'_{mn}\g^{mn}\eta_+
-\beta^*\frac{e^{\phi/4}}{192}G_{mnpq}\g^{mnpq}\eta_+
\la{pevte}\\
0&=\frac{1}{2}\beta\partial_m\phi \g^m  \eta_- +
\beta\frac{e^{-\phi/2}}{24}H_{mnp}\g^{mnp}\eta_-
+\alpha\frac{5me^{5\phi/4}}{4}\eta_+\nn\\
& +i\alpha f\frac{e^{\phi/4}}{8}\eta_+
-\alpha\frac{3me^{3\phi/4}}{8}B'_{mn}\g^{mn}\eta_+
+\alpha\frac{e^{\phi/4}}{192}G_{mnpq}\g^{mnpq}\eta_+
~.
\la{e3i}
\end{align}

\subsection{$SU(3)$ structure and tensor decomposition}

The existence of the spinor $\eta$ allows us to define the bilinears
\be
J_{mn}:=i\eta^+_-\g_{mn}\eta_- = -i\eta^+_+\g_{mn}\eta_+
\la{j}
\end{equation}
\be
\Omega_{mnp}:=\eta_-^+\g_{mnp}\eta_+; ~~~~~~ 
\Omega^*_{mnp}=-\eta_+^+\g_{mnp}\eta_-~.
\la{o}
\end{equation}
Note that $J_{mn}$ thus defined is real and 
$\Omega$ ($\Omega^*$) is imaginary (anti-) self-dual, as
can be seen from (\ref{hodge})
\be
\Omega_{mnp}=\frac{i}{6}\sqrt{g_6}~\varepsilon_{mnpijk}\Omega^{ijk}~.
\end{equation}
We choose to  normalize 
\be
\eta_+^+\eta_+=\eta_-^+\eta_-=1~.
\la{n}
\end{equation}
Using (\ref{fierz}) one can prove 
that $J$, $\Omega$ satisfy
\be
J_m{}^nJ_n{}^p=-\delta_m{}^p
\end{equation}
\be
(\Pi^+)_m{}^n\Omega_{npq}=\Omega_{mpq}; ~~~~~~ 
(\Pi^-)_m{}^n\Omega_{npq}=0  ~,
\end{equation}
where 
\be
(\Pi^\pm)_m{}^n:=\frac{1}{2}(\delta_m{}^n\mp i J_m{}^n)
\la{projectors}
\end{equation}
are the projection operators onto the holomorphic/antiholomorphic parts.
In other words, $J$ defines an almost complex structure 
with respect to which $\Omega$ is $(3,0)$.
Moreover (using (\ref{fierz}) again) it follows that
\beal
\Omega\wedge J&=0\nn\\
\Omega\wedge\Omega^*&=\frac{4i}{3}J^3~.
\end{align}
Therefore $J$, $\Omega$, completely specify an $SU(3)$ structure on $X_6$. 
Some further useful identities are given in appendix B.

In the case of  
a manifold $X_6$ of 
$SU(3)$ structure, 
the intrinsic torsion decomposes into five modules (torsion classes) 
${\cal W}_1\dots {\cal W}_5$. These also appear in the 
$SU(3)$ decomposition of the exterior derivative of $J$, $\Omega$. 
Intuitively this should be clear since the intrinsic torsion parameterizes the 
failure of the manifold to be of special holonomy, which can also 
be thought of as the failure of the closure of $J$, $\Omega$. 
More specifically we have
\beal
dJ&=-\frac{3}{2}Im(\cw_1\Omega^*)+\cw_4\wedge J+\cw_3\nn\\
d\Omega&= \cw_1 J\wedge J+\cw_2 \wedge J+\cw_5^*\wedge \Omega ~.
\label{torsionclasses}
\end{align}
The classes $\cw_1$, $\cw_2$ can be decomposed further into real and imaginary parts 
$\cw^{\pm}_1$, $\cw^\pm_2$.

As a final ingredient before we proceed to the analysis of the next 
section, we will need the decomposition of the form fields 
with respect to the reduced structure group $SU(3)$. 
Using the projectors (\ref{projectors}) we can 
decompose the tensors $B'$, $H$, $G$ in terms of irreducible 
representations. Explicitly (we henceforth drop the prime on $B$),
\beal
B_{mn}=\frac{1}{16}\Omega^*_{mn}{}^sB^{(1,0)}_{s}
+\frac{1}{16}\Omega_{mn}{}^sB^{(0,1)}_{s}
+(\tilde{B}_{mn}+\frac{1}{6}J_{mn}B^{(0)})~,
\la{bexp}
\end{align}
where $B^{(1,1)}_{mn}$ has been 
further decomposed into traceless ($\tilde{B}$) 
and trace (${B}^{(0)}$) parts. 
The normalization above has been chosen so that
\beal
B^{(0)}&=B_{mn}J^{mn}\nn\\
B^{(1,0)}_m&=\Omega_m{}^{np}B_{np}~.
\end{align}
In terms of $SU(3)$ representations we have,
\be
B^{(0)}\sim {\bf 1}; 
~~~~~ B^{(1,0)}\sim {\bf 3}; ~~~~~ 
B^{(0,1)}\sim {\bf \bar{3}}; ~~~~~ 
\tilde{B}\sim {\bf {8}}~.
\end{equation}
Note that the tracelessness of $\tilde{B}$ is 
equivalent to the primitivity condition
\be
J\wedge J\wedge \tilde{B}=0~.
\label{primitive}
\end{equation}
Similarly for the $H$ field we expand,
\be
H_{mnp}=\frac{1}{48}\Omega_{mnp}H^{(0)}
+(\tilde{H}_{mnp}^{(2,1)}+\frac{3}{4}H^{(1,0)}_{[m}J_{np]}  ) 
+{\rm c.c.}~,
\la{hexp}
\end{equation}
where
\beal
H^{(0)}&=\Omega^{*mnp}H_{mnp}\nn\\
H^{(1,0)}_m&=(\Pi^+)_m{}^sH_{snp}J^{np}~
\end{align}
and
\be
\tilde{H}^{(2,1)}\sim {\bf {6}}; ~~~~~ \tilde{H}^{(1,2)}\sim {\bf \bar{6}}~.
\end{equation}
Finally, for the four-form $G$ we have,
\be
G_{mnpq}=\frac{1}{12}G^{(1,0)}_{[m}\Omega^{*}_{npq]}
+\frac{1}{12}G^{(0,1)}_{[m}\Omega_{npq]}
+(3\tilde{G}_{[mn}J_{pq]} 
+\frac{1}{8}G^{(0)}J_{[mn}J_{pq]} )~,
\la{gexp}
\end{equation}
where
\beal
G^{(0)}&=G_{mnpq}J^{mn}J^{pq}\nn\\
G^{(1,0)}_m&=\Omega^{npq}G_{mnpq}\nn\\
\tilde{G}_{mn}
&=2(\Pi^+)_m{}^s(\Pi^-)_n{}^tG_{stpq}J^{pq}-\frac{1}{6}J_{mn}G^{(0)}
~.
\end{align}
Note that the scalars $B^{(0)}$, $G^{(0)}$ are real whereas 
$H^{(0)}$ is complex.

\section{Analysis}
\label{lisp}

We will start by examining the content of the supersymmetry equations 
(\ref{eva}-\ref{e3i}), the Bianchi identities (\ref{prwtn},\ref{bianchi1}) 
and the form equations of motion (\ref{heq},\ref{geq}).  
In section \ref{thetorsions} we read off  the torsion classes 
of the $SU(3)$ manifold $X_6$. Some special 
cases of our solutions are examined in \ref{special}.
Equations (\ref{a1},\ref{a2}) of appendix 
\ref{appx} will be 
very useful in the following.

\subsection{Supersymmetry}
\label{analysis}

We are now ready to analyze the content of equations (\ref{eva}-\ref{e3i}). 
Our strategy will be to perform all possible contractions with 
$\eta_\pm^+\g^{(n)}$, as is made clear by the following

{\bf Lemma:} For $\chi$, $\e$ constant spinors in $\mathbb{R}^{6}$, where $\chi$ 
is non-vanishing, 
$$
\e=0 \Longleftrightarrow 
\chi\g^{(n)}\e=0, ~~ n=0,\dots , 3~.
$$
Proof: First note that 
\be
\e^\alpha=0 \Longleftrightarrow 
\xi^\alpha C_{\alpha\beta} \e^\beta=0, ~~ \forall \xi~,
\la{3.1}
\end{equation}
where $\e$, $\xi$ are spinors in $\mathbb{R}^{6}$.  
Clearly, if $\e=0$ it follows 
that $\xi^\alpha C_{\alpha\beta} \e^\beta=0$. Conversely, if $\e\neq 0$ we 
can assume without loss of generality that $\e^{\alpha=1}\neq 0$ and 
$C_{\alpha=2~\beta=1}\neq 0$ 
(it cannot be that $C_{\alpha 1}=0,~~\forall \alpha$, 
as this would imply $det(C)=0$ whereas $C$ is 
in fact unitary). It follows that 
$\xi^\alpha C_{\alpha\beta} \e^\beta\neq 0$, for $\xi^\alpha=\delta_2^\alpha$.
Moreover it holds that if $\chi$ is a non-vanishing spinor, 
for any $\xi$ there exist constants $\{\phi^{(n)}_{a_1\dots a_n}\}$ such that
\be
\xi=\sum_{n=0}^3 \phi^{(n)}_{a_1\dots a_n}\g^{a_1\dots a_n}\chi~.
\la{3.2}
\end{equation}
This follows from the fact that the Gamma matrices 
$\{ \g^{(n)} , n=0,\dots 3\}$ 
generate $Gl(4,\mathbb{C})$ which acts transitively 
on $\mathbb{C}^4-\{0\}\ni \chi$. 
For globally-defined spinors
on a manifold $X_6$, as 
is the case at hand, the above can be generalized to arbitrary 
points on the tangent bundle. In this case  
$\{\phi^{(n)}_{a_1\dots a_n}\}$ become 
forms on $X_6$. The lemma follows immediately from (\ref{3.1}, \ref{3.2}).

\subsection*{The {\bf 1}}

We first note that multiplying (\ref{pevte}) by $\beta$ (\ref{e3i}) by $\alpha^*$ 
adding them together and separating real and imaginary parts, we obtain
\begin{align}
0&=(|\alpha|^2-|\beta|^2)(\frac{5me^{5\phi/4}}{4}-\frac{e^\phi}{64}G^{(0)}  )\nn\\
0&=mB^{(0)}-\frac{fe^{-\phi/2}}{3}  ~,
\label{above}
\end{align}
where we noted that $|\alpha|^2+|\beta|^2 >0$. We therefore distinguish two cases:

{\it Case 1: $|\alpha|\neq |\beta|$}.

The first of equations  (\ref{above}) implies
\begin{equation}
G^{(0)}=80me^{\phi}~.
\end{equation}
Substituting this back to (\ref{pevte}), (\ref{e3i}) we obtain
\begin{equation}
H^{(0)}=0~.
\end{equation}
Similarly, multiplying (\ref{tria}) by $\beta$ (\ref{tessera}) by $\alpha$ 
subtracting one from the other and separating 
real and imaginary parts, we obtain
\begin{align}
m&=0\nn\\
F^{(0)}&=0~,
\end{align}
where in the second line we have taken the massless limit (\ref{masslesslimit}). 
Finally, 
plugging the above back to (\ref{tria}), (\ref{tessera}) implies 
\begin{equation}
W=0~.
\end{equation}
Hence, this case reduces to compactification to $\mathbb{R}^{1,3}$. This 
was analyzed in detail in \cite{getala}, \cite{getalb}, \cite{dp}  and will not 
concern us further here.

{\it Case 2: $|\alpha| =|\beta|$}.

Without loss of generality, we can choose the phase of the internal 
spinor $\eta$ so that 
\begin{equation}
\alpha=\beta\neq 0~.
\label{fix}
\end{equation}
In the following it will be useful to add and subtract 
equations (\ref{eva},\ref{duo}), taking 
(\ref{fix}) into account, to obtain
\beal
0&=\nabla_m\eta_+
+\frac{1}{2}\partial_mln|\alpha|^2\eta_+
+3if\frac{e^{\phi/4}}{32}\g_m\eta_-
+\frac{me^{3\phi/4}}{32}B'_{np}
(\g_m{}^{np}-14\delta_m{}^n\g^p)\eta_-
\la{evaprime}\\
0&=
+\frac{1}{2}\partial_m ln\Big(\frac{\alpha}{\alpha^*}\Big)\eta_+
+\frac{e^{-\phi/2}}{96}H_{npq}
(\g_m{}^{npq}-9\delta_m{}^n\g^{pq})\eta_+
-\frac{me^{5\phi/4}}{16}\g_m\eta_-\nn\\
&+\frac{e^{\phi/4}}{256}G_{npqr}(\g_m{}^{npqr}-\frac{20}{3}\delta_m{}^n
\g^{pqr})\eta_-~,
\la{duoprime}
\end{align}
In this case the system of equations (\ref{tria}-\ref{e3i}), (\ref{duoprime}) 
can be solved to give
\beal
mB^{(0)}&=\frac{1}{3}fe^{-\phi/2}\nn\\
H^{(0)}&=\frac{96}{5}me^{7\phi/4}\nn\\
G^{(0)}&=\frac{144}{5}me^\phi\nn\\
W&=\Delta\Big( \frac{\alpha}{|\alpha|}\Big)^{-2}
(-\frac{1}{5}me^{5\phi/4}+\frac{i}{6}fe^{\phi/4})~.
\label{singletsolution}
\end{align}
\subsection*{The  3}

The solution of equations (\ref{tria}-\ref{e3i})  
reads
\beal
\partial^{(1,0)}_m\phi&=\frac{3}{8}me^{3\phi/4}B^{(1,0)}_m\nn\\
H^{(1,0)}_m&=0\nn\\
G^{(1,0)}_m&=0\nn\\
\Delta&=constant\times e^{-\phi/12}~,
\end{align}
where we have defined
\be
\partial^{(1,0)}_m:=(\Pi^+)_m{}^n\partial_n~.
\end{equation}
Equation (\ref{duoprime}) then implies 
\be
Arg(\alpha)=constant~,
\end{equation}
which, taking (\ref{singletsolution}) into account  and the fact 
that $W$ is a constant,  implies the following two cases:
\beal
\phi,~f &=constant\nn\\
B^{(1,0)}&=0  
\end{align}
and $m\neq 0$, or,
\beal
f&=constant\times e^{-\phi/6}\nn\\
Arg(\alpha)&=\frac{\pi}{4}  
\end{align}
and $m=0$. The latter case can be seen to reduce to four-dimensional 
Minkowski space once the Bianchi identities and the equations of motion are imposed, 
and will not concern us further. 

\subsection*{The  6}

This representation drops out of equations (\ref{tria}-\ref{e3i}). 
Equation (\ref{duoprime}) implies
\be
\tilde{H}^{(1,2)}=0~.
\end{equation}

\subsection*{The  8}

As in the previous case, this representation drops out of equations (\ref{tria}-\ref{e3i}). 
Equation (\ref{duoprime}) implies
\beal
\tilde{G}=0~.
\end{align}
To summarize our results so far, 
the solution to equations (\ref{tria}-\ref{e3i},\ref{duoprime}) reads 
in form notation

\begin{center}
\fbox{\parbox{9.6cm}{
%
\beal
mB&=\frac{f}{18}e^{-\phi/2}J+m\tilde{B}\nn\\
H&=\frac{4m}{5}e^{7\phi/4}Re(\Omega)\nn\\
G&=fdVol_4+\frac{3m}{5}e^\phi J\wedge J\nn\\
W&=\Delta\Big( \frac{\alpha}{|\alpha|}\Big)^{-2}
(-\frac{1}{5}me^{5\phi/4}+\frac{i}{6}fe^{\phi/4})\nn\\
\phi,&~\Delta,~f,~Arg(\alpha)=constant ~.\nn
\end{align}
}}
\end{center}
\be\label{solution}\end{equation}
In the above we have denoted by $dVol_4$ the volume element 
of $AdS_4$ in the warped metric.

\subsection*{The $SU(3)$ structure}

Plugging equation (\ref{evaprime}) 
into the definitions (\ref{j},\ref{o}) we find
\beal
\nabla_mJ_{kl}&= -J_{kl}\partial_mln|\alpha|^2+\frac{2}{9}fe^{\phi/4}
Re(\Omega_{mkl})+me^{3\phi/4}Im(\Omega_{kl}{}^s)\tilde{B}_{ms}  \nn\\
\nabla_m\Omega_{klt}&=-\Omega_{klt}\partial_mln|\alpha|^2 
+6ime^{3\phi/4}J_{[kl}(\Pi^+)_{t]}{}^s\tilde{B}_{sm}
-\frac{4}{3}fe^{\phi/4}J_{[kl}(\Pi^+)_{t]m}  ~.
\label{nablajo}
\end{align}
By antisymmetrizing in all indices we obtain
\be
dJ=-J\wedge dln|\alpha|^2+\frac{2}{3}fe^{\phi/4}Re(\Omega)
\label{dj}
\end{equation}
and
\be
d\Omega=-\Omega\wedge dln|\alpha|^2-2ime^{3\phi/4}J\wedge\tilde{B}
-\frac{4i}{9}fe^{\phi/4}J\wedge J   ~.
\label{do}
\end{equation}

\subsection{Equations-of-motion and Bianchi identities}

Taking (\ref{solution}) into account, 
the Bianchi identity (\ref{bianchi1}) for the $H$ field 
implies
\be
dRe(\Omega)=0~,
\label{mkkm}
\end{equation}
which is satisfied iff 
\begin{center}
\fbox{\parbox{3.5cm}{
%
%
\be
|\alpha|=constant~,\nn
\end{equation}
%
%
}}
\end{center}
\be\label{kirkuk}\end{equation}
as can be seen from equation (\ref{do}). The Bianchi identity (\ref{bianchi1}) for the $G$ field 
can be seen to be satisfied automatically by noting that 
$\tilde{B}\wedge\Omega=0$ and 
$dJ\wedge J\propto Re(\Omega)\wedge J=0$. 
In the latter we have taken (\ref{dj},\ref{kirkuk}) into account. 
Moreover, using (\ref{solution}) we see that equation (\ref{prwtn}) 
implies
\begin{center}
\fbox{\parbox{7cm}{
%
%
\be
md\tilde{B}=\frac{e^{-\phi/4}}{27}(\frac{108m^2}{5}e^{2\phi}-f^2 )Re(\Omega) ~,\nn
\end{equation}
}}
\end{center}
\be\label{db}\end{equation}
from which it follows that
\be
d*\tilde{B}=0~.
\label{kra} 
\end{equation}
In deriving (\ref{kra}) we noted that $J\wedge d\tilde{B}=d(J\wedge \tilde{B})=
d*(\tilde{B}\wedge dVol_4)$, 
as can be seen from (\ref{dj}, \ref{kokoriko}). It follows from (\ref{kra}) 
and the Hodge 
decomposition theorem that $*\tilde{B}$ is harmonic up to an exact form: 
$*\tilde{B}=d\chi+Harm$, i.e. $d\tilde{B}=-d*d\chi$, for 
some globally-defined seven-form $\chi$. We can see that this is indeed 
the case: using (\ref{db}) it can be shown that 
\beal
*d*dIm(\Omega)&=\frac{2}{3}e^{\phi/2}(\frac{12m^2}{5}e^{2\phi}-f^2)
Im(\Omega)\nn\\
d*d*Re(\Omega)&=\frac{2}{3}e^{\phi/2}(\frac{12m^2}{5}e^{2\phi}-f^2)
Re(\Omega)
\label{extracondition}
\end{align}
and therefore $d\tilde{B}\propto Re(\Omega)\propto 
d*d*Re(\Omega)\propto 
d*d Im(\Omega)\wedge dVol_4$. Similar equations have appeared 
in the mathematical literature in \cite{cswann}.

Note that 
(\ref{kirkuk},\ref{kra}) are equivalent to the consistency condition $d(d\Omega)=0$. 
The corresponding condition for the almost complex structure, $d(dJ)=0$, is 
automatically satisfied. However, equation (\ref{db})  
is a consequence of supersymmetry and the Bianchi identities,  
and has to be imposed as an 
extra condition on the $SU(3)$ structure. We will examine a class 
of examples with this property in section \ref{examples}.

Starting from $d(\Omega\wedge \tilde{B})=0$, taking 
(\ref{do},\ref{db}) into account, one arrives at
\beal
m^2|\tilde{B}|^2&=-\frac{4e^{-\phi}}{27}(\frac{108m^2}{5}e^{2\phi}-f^2)~,\nn\\
f^2&\geq  \frac{108m^2}{5}e^{2\phi}~.
\label{bmeasure}
\end{align}
Using (\ref{kokoriko}), it can then be seen that 
the form equations (\ref{heq},\ref{geq}) are 
satisfied with no further restrictions on the fields. Although not necessary,  
as was argued in section \ref{kiki}, 
we have checked that the dilaton equation is also satisfied.

\subsection{Torsion classes}
\label{thetorsions}

Equations (\ref{dj},\ref{do}) 
can be used to read off 
the torsion classes of the internal manifold $X_6$ by 
comparing with (\ref{torsionclasses}) (taking (\ref{kirkuk}) into account):
\begin{center}
\fbox{\parbox{5.5cm}{
%
%
%
\beal
\cw^+_1&=0\nn\\
{\cal W}^-_1&=-\frac{4i}{9}fe^{\phi/4}\nn\\
\cw_2^+&=0\nn\\
{\cal W}^-_2&=-2ime^{3\phi/4}\tilde{B}\nn\\
{\cal W}_3&=0\nn\\
{\cal W}_4&=0\nn\\
{\cal W}_5&=0 ~.\nn
\end{align}
%
}}
\end{center}
\be\label{torsions}\end{equation}
Note that $X_6$ belongs to the class of 
half-flat manifolds. As mentioned in the introduction, the latter are defined by the property 
$\cw_1^+=\cw_2^+=\cw_4=\cw_5=0$.

In conclusion, type IIA, ${\cal N}=1$ compactifications on  
$AdS_4\times X_6$ are given by (\ref{solution},\ref{kirkuk}), where 
the internal manifold $X_6$ has $SU(3)$ structure with torsion classes given by (\ref{torsions}) 
and ${\cal W}^-_2$ is further restricted by (\ref{db}).

\subsection{Special cases}
\label{special}

In the limit where
\be
f^2=\frac{108m^2}{5}e^{2\phi}~,
\end{equation}
it follows from (\ref{bmeasure}) that $\tilde{B}$ and $\cw_2$ vanish.  
Then all torsion classes vanish except 
for $\cw_1^-$, and $X_6$ is further restricted to be {\it nearly-K\"{a}hler}. This case was 
analyzed in detail in \cite{bca, bc}.

The massless limit should be taken with care, as was explained in section \ref{kiki}.
In this case the solution reduces to
\beal
F&=\frac{1}{9}f e^{-\phi/2}J+\tilde{F}\nn\\
H&=0\nn\\
G&=fdVol_4\nn\\
W&=\frac{i}{6}\Delta\Big( \frac{\alpha}{|\alpha|}\Big)^{-2}fe^{\phi/4}\nn\\
\phi,&~\Delta,~f,~\alpha=constant ~
\end{align}
with
\beal
d\tilde{F}=-\frac{2}{27} f^2 e^{-\phi/4}Re(\Omega)~.
\end{align}
All torsion classes are zero except for $\cw_1^-$, $\cw_2^-$ which are given by
\beal
{\cal W}^-_1&=-\frac{4i}{9}fe^{\phi/4}\nn\\
{\cal W}^-_2&=-ie^{3\phi/4}\tilde{F}~.
\end{align}
This solution can be lifted to eleven dimensions, leading 
to a seven-dimensional internal manifold which is a (twisted) 
circle fibration over a six-dimensional half-flat base. 
Compactifications of eleven dimensional 
supergravity to $AdS_4$ in which the seven-dimensional internal space is a product 
of a circle and 
a six-dimensional nearly K\"{a}hler manifold, were considered in \cite{l}. 
However no well-defined solutions were obtained in this reference. 
In fact, the results of the present paper imply that no such solutions exist. 
This can be seen as follows: in order for the manifold $X_6$ to be nearly K\"{a}hler 
we would have to have $f=0$. Note that taking the $f\rightarrow 0$ limit 
naively appears to lead to a compactification 
on $\mathbb{R}^{1,3}\times X_6$ with only the primitive 
part of the $F$-flux turned on.   
However, as we can see from (\ref{bmeasure}), $F$ has to vanish 
and the internal manifold reduces to a Calabi-Yau.

\section{Examples}
\label{examples}

\subsection{$T^2$ over K3}
\label{t2overk3}

We now construct a class of examples of six dimensional 
manifolds $X_6$ with the property that their intrinsic torsion 
is contained in $\cw_1^-\oplus\cw_2^-$ and, in addition, 
the exterior derivative of $\cw_2^-$ is proportional 
to $Re(\Omega^{(3,0)} )$ 
\footnote{In this subsection we will write $\Omega^{(3,0)}$ instead 
of simply $\Omega$, to distinguish from the 
holomorphic two-form $\Omega^{(2,0)}$ defined in the following.}. 
Our starting point is the work of Goldstein and Prokushkin 
\cite{gopr}. These authors have shown that 
six-dimensional manifolds with $SU(3)$ structure 
can be constructed as $T^2$ fibrations over Hermitian four-dimensional 
manifolds ($X_4$). The metric on the total space is then of the form
\beal
g_{base}+(dx+a)^2+(dy+b)^2~,
\label{gt}
\end{align}
where $g_{base}$ is the metric on the base $X_4$ and $a$, $b$ are local 
one-forms on $X_4$. Moreover $a$, $b$ satisfy $da=\omega_P$, $db=\omega_Q$, with
\beal
\frac{[\omega_P]}{2\pi},~\frac{[\omega_Q]}{2\pi}\in H^2(X_4,\mathbb{Z})~.
\label{quantization}
\end{align}
The complex $(3,0)$ form and the almost complex structure on the total space are
given by
\beal
\Omega^{(3,0)}&=\{(dx+a)+i(dy+b)  \}\wedge \Omega^{(2,0)} 
\label{toto}
\end{align}
and
\beal
J=\omega+(dx+a)\wedge(dy+b)~,
\label{totj}
\end{align}
where $\Omega^{(2,0)}$, $\omega$ are the holomorphic $(2,0)$ form and 
the hermitian $(1,1)$ form on the base\footnote{If the map $\pi : X_6\mapsto X_6/ T^2 \simeq X_4$ 
defines the fibration, we can extend $\Omega^{(2,0)}$, $\omega$, $a$, $b$ 
from $X_4$ to the total space $X_6$ by using $\pi^*$.}
 respectively. In the case we are considering $X_4$ is a Calabi-Yau two-fold (i.e. a K3 surface) and 
therefore $\Omega^{(2,0)}$, $\omega$ are closed. 

The two-forms $\omega_P$, $\omega_Q$ should have no component 
in $\Lambda^{0,2}T^*X_4$ in order for the total space $X_6$ to be complex. 
However, as was noted in \cite{gopr}, this condition can be relaxed. In fact, for our 
purposes we will take $\omega_P$, $\omega_Q$ to be purely of type $(2,0)\oplus(0,2)$ 
on the base. Namely we take
\beal
\omega_P&=-\frac{3i}{2}\cw_1^- Im(\Omega^{(2,0)})\nn\\
\omega_Q&=-\frac{3i}{2}\cw_1^- Re(\Omega^{(2,0)})~,
\label{wpq}
\end{align}
where $\cw_1^-$ is an imaginary constant, which should be 
quantized. We can see this as follows:  
let $\Gamma_{3,19}$ be the even, self-dual lattice of integral cohomology, where
the following identifications are understood,
\beal
\Gamma_{3,19}\simeq H^2(X_4,\mathbb{Z}) \subset H^2(X_4,\mathbb{R})~.
\end{align}
Note that $\Omega^{(2,0)} \in H^2(X_4,\mathbb{C})\simeq H^2(X_4,\mathbb{R})\oplus H^2(X_4,\mathbb{R})$. 
Let us define $x:=Re(\Omega^{(2,0)})$, $y:=Im(\Omega^{(2,0)})$ so that $x,~y \in H^2(X_4,\mathbb{R})$. 
For $u, ~v \in H^2(X_4,\mathbb{R})$ we can define an inner product by
\beal
u\cdot v :=\int_{X_4}u\wedge v~.
\end{align}
We have, 
\beal
(x\cdot x+y\cdot y)=\int_{X_4}\Omega^{(2,0)}\wedge\Omega^{(0,2)}= 4 Vol(X_4)~.
\end{align}
Moreover, it follows from $\Omega^{(2,0)}\wedge \Omega^{(2,0)}=0$ 
that $x\cdot y=0$, $x\cdot x=y\cdot y$, and therefore 
\beal
\frac{1}{Vol(X_4)} x\cdot x =2~.
\end{align}
I.e. $x,y$ have the same length and are orthogonal to each other. Let
$\Sigma$ be the two-plane defined by $x$, $y$.
Changing the complex structure on $X_4$ while keeping $\Gamma_{3,19}$ fixed
causes $\Sigma$ to rotate, spanning the entire space of two-planes in $H^2(X_4,\mathbb{R})$
(see for example \cite{a}). By choosing an appropriate complex
structure on $X_4$, we may arrange so that $x/\sqrt{Vol(X_4)}, ~y/\sqrt{Vol(X_4)} 
\in \Gamma_{3,19}$, as
can be seen from the explicit form of the lattice.
It follows from (\ref{wpq}) that in order
for (\ref{quantization}) to hold, $\cw_1^-$ has to be quantized.

From  (\ref{toto},\ref{totj},\ref{wpq}) we can compute the exterior 
derivatives of $\Omega^{(3,0)}$, $J$,
\beal
dJ&=\frac{3i}{2}\cw_1^- Re(\Omega^{(3,0)}) \nn\\
d\Omega^{(3,0)}&=\frac{3}{2}\cw_1^- \Omega^{(2,0)}\wedge\Omega^{(0,2)}
~.
\end{align}
Moreover, by noting that $\Omega^{(2,0)}\wedge\Omega^{(0,2)}=2\omega\wedge\omega$, 
we can see that the last line can be written as
\beal
d\Omega^{(3,0)}=\cw_1^- J\wedge J+\cw_2^- \wedge J 
~,
\end{align}
where 
\beal
\cw_2^- :=2\cw_1^- \{\omega-2(dx+a)\wedge(dy+b)\}~.
\label{quirk}
\end{align}
Note that $\cw_2^-$ satisfies the primitivity condition $J\wedge J\wedge \cw_2^-=0$. 
By comparing with (\ref{torsionclasses}) we conclude that the intrinsic torsion 
of $X_6$ is entirely within $\cw_1^-\oplus\cw_2^-$. In addition, from (\ref{quirk}) 
we see that the exterior derivative of $\cw_2^-$ is proportional 
to $Re(\Omega^{(3,0)})$  as promised: 
\beal
d\cw_2^-=-6i(\cw_1^-)^2 Re(\Omega^{(3,0)})~.
\label{dw2}
\end{align}
However as we can see from (\ref{db},\ref{torsions}), 
there are no values of $f$, $m$, other than $f=m=0$, that can 
fit with the above equation. In other words, although the example examined in this section
satisfies (\ref{i1}, \ref{i2}), it does not correctly reproduce the proportionality 
constant in equation (\ref{db}) except for the rather special case 
where all fluxes vanish and the internal 
manifold reduces to a Calabi-Yau threefold.

\subsection{Iwasawa manifold}
\label{iwasawa}

The following example, given in \cite{cs},  of dynamic half-flat $SU(3)$ 
structure is based on the Iwasawa manifold ${\cal M}$. 
Consider the following basis of one-forms on ${\cal M}$:
\begin{alignat}{2}
de^5&=-e^{14}-e^{23}\nn\\
de^6&=-e^{13}-e^{42}   \nn\\ 
de^i&=0,   \qquad\qquad\qquad\qquad i=1,2,3,4
~,
\label{de}
\end{alignat}
where we use the notation $e^{ij}:=e^i\wedge e^j$. Then for all 
$t\in \mathbb{R}^+$ the following defines 
a half-flat $SU(3)$ structure
\beal
M^{2}J&= t^2(e^{12}+e^{34})+t^{-2}e^{56} \nn\\
M^{3}\Omega&=t(e^1+ie^2)\wedge(e^3+ie^4)\wedge(e^5+ie^6) ~,
\end{align}
compatible with the metric 
\beal
M^{2}g= t^2\sum_{i=1}^4e^i\otimes e^i+t^{-2}\sum_{i=5}^6 e^i\otimes e^i ~.
\label{metric}
\end{align}
We have introduced a mass scale $M$ so that the einbeine $e^i$ are dimensionless. 
It is then straightforward verify that 
\beal
dJ&=\frac{3i}{2}\cw_1^-Re(\Omega)\nn\\
d\Omega&=~\cw_1^- J\wedge J +\cw_2^- \wedge J,
\end{align}
where
\beal
M^{-1}\cw^-_1&:= -\frac{2i}{3}t^{-3}\nn\\
M\cw_2^-&:= \frac{8i}{3}(
-\frac{1}{2}t^{-1}e^{12}
-\frac{1}{2}t^{-1}e^{34}
+t^{-5}e^{56})   ~.
\end{align}
By comparing with (\ref{torsionclasses}) we see that the intrinsic torsion 
is contained in $\cw_1^-\oplus\cw_2^-$. 
Note that $\cw_2^-$ satisfies the primitivity condition $\cw_2^-\wedge J\wedge J=0$, as it should. 
Moreover we find
\beal
d\cw_2^- =-6i(\cw_1^-)^2 Re(\Omega) ~,
\end{align}
which is the same as equation (\ref{dw2}) of the previous example! 
However, this is not a coincidence as the example based on the Iwasawa 
manifold is in fact a special case of the $T^2$ over $K3$ fibration considered 
in \ref{t2overk3}. This can be seen as follows: equations (\ref{de}) allow 
us to express the vielbeine in terms of coordinates $x^i$ such that
\footnote{In the following we set $t, ~M=1$ for simplicity.}
\beal
e^5&=dx^5-x^1dx^4+x^3dx^2\nn\\
e^6&=dx^6-x^1dx^3-x^4dx^2\nn\\
e^i&=dx^i, ~~~~~~~~~~~~~~~~~~~~~~~~~~~~~~~~~~~~i=1,2,3,4
~.
\end{align}
The coordinates $x^i, ~~i=1\dots 4$, parameterize a $T^4$ base on which 
we can define a hermitian (in fact K\"{a}hler) $(1,1)$ form and a holomorphic 
(2,0) form in analogy with the previous section:
\beal
\omega&=\frac{i}{2}(dz^1\wedge d\bar{z}^1 + dz^2\wedge d\bar{z}^2)\nn\\
\Omega^{(2,0)}&=dz^1\wedge dz^2~,
\end{align}
where $dz^1:=dx^1+idx^2$, $dz^2:=dx^3+idx^4$. 
Equation  (\ref{metric}) can be written as 
\beal
g=\sum_{i=1}^4(dx^i)^2+(dx^5+a)^2+(dx^6+b)^2~,
\end{align}
where $a:=-x^1dx^4+x^3dx^2$, $b:=-x^1dx^3-x^4dx^2$. 
I.e. the metric is of the form (\ref{gt}) and, 
moreover, it can be seen that (\ref{wpq}, \ref{dw2})  are satisfied  
for $\cw_1^-=-2i/3$. In other words, the example based on the Iwasawa manifold 
is a degenerate instance of the $T^2$ over $K3$ fibration presented 
in \ref{t2overk3}, whereby the $K3$ base is replaced  
by a $T^4$.

\section*{Acknowledgments}

We would like to thank Claus Jeschek for useful discussions.

\appendix
\section{Spinor conventions}

In all dimensions the Gamma matrices are taken to obey
\be
(\G^M)^+=\G^0\G^M\G^0~,
\end{equation}
where the Minkowski metric is mostly plus. Antisymmetric products of 
Gamma matrices are defined by
\be
\G^{(n)}_{M_1\dots M_n}:=\G_{[M_1}\dots\G_{M_n]}~.
\end{equation}

\subsection{Spinors in $D=6$}

The charge conjugation matrix in six Euclidean dimensions satisfies
\be
C^{Tr}=C; ~~~~~~ (C\g^m)^{Tr}=-C\g^m~.
\end{equation}
The fundamental (4-dimensional, chiral) spinor representation 
$\eta_+$ is complex and we define $\eta_-$ by
\be
\eta^+_+=\eta_-^{Tr}C~,
\end{equation}
which also implies
\be
\eta^+_-=\eta_+^{Tr}C~.
\end{equation}
A useful formula which follows from the above is
\be
{(\g^{(n)}\eta_{\pm})}^*=(-)^nC\g^{(n)}\eta_{\mp}~.
\end{equation}
The chirality matrix defined by
\be
\g_7:=-i\g_1\dots\g_6; ~~~~~~\g_7^2=1 ~,
\end{equation}
can be used to express the Hodge-dual of an antisymmetric product of 
gamma-matrices
\be
i\g^{(n)}=(-)^{\frac{1}{2}k(k-1)}*\g^{(D-n)}\g_7~.
\la{hodge}
\end{equation}
Fierz rearrangement follows from
\be
\chi_{\pm}^\alpha\psi^\beta =\frac{1}{4}\phi^\pm(P_\pm C^{-1})^{\alpha\beta}
-\frac{1}{4}
 \phi^\pm_m(P_\pm \g^mC^{-1})^{\alpha\beta}+
\frac{1}{8} \phi^\pm_{mn}(P_\pm \g^{mn}C^{-1})^{\alpha\beta}-
\frac{1}{48}
 \phi^\pm_{mnp}(P_\pm \g^{mnp}C^{-1})^{\alpha\beta}~,
\end{equation}
where
\be
\phi_{m_1\dots m_k}:=\chi_{\pm}^{Tr}C\g_{m_1\dots m_k}\psi~
\end{equation}
and 
\be
P_\pm :=\frac{1}{2}(1\pm\g_7)~.
\end{equation}
Note that $\phi_{mnp}^{+}$ ($\phi_{mnp}^{-}$) 
is imaginary (anti-) self-dual, as 
follows from (\ref{hodge}). In particular, using definitions 
(\ref{j},\ref{o}) 
we find
\beal
\eta_-^\alpha\eta_+^\beta &=\frac{1}{4}(P_- C^{-1})^{\alpha\beta}
+\frac{i}{8}J_{mn}(P_- \g^{mn}C^{-1})^{\alpha\beta}\nn\\
\eta_+^\alpha\eta_+^\beta &=-\frac{1}{48}
\Om_{mnp}(P_+ \g^{mnp}C^{-1})^{\alpha\beta}\nn\\
\eta_-^\alpha\eta_-^\beta &=\frac{1}{48}
\Om^*_{mnp}(P_- \g^{mnp}C^{-1})^{\alpha\beta}~.
\la{fierz}
\end{align}

\subsection{Spinors in $D=1+3$}

The charge conjugation matrix in $1+3$ dimensions satisfies
\be
C^{Tr}=-C; ~~~~~~ (C\g^\mu)^{Tr}=-C\g^\mu~.
\end{equation}
The fundamental (2-dimensional, chiral) spinor representation 
$\theta_+$ is complex and we define $\theta_-$ by
\be
\overline{\theta}_+=\theta_-^{Tr}C~,
\end{equation}
which also implies
\be
\overline{\theta}_-=-\theta_+^{Tr}C~,
\end{equation}
where
\be
\overline{\theta}:=\theta^+\g^0~.
\end{equation}
A useful formula which follows from the above is
\be
{(\g^{(n)}\theta_{\pm})}^*=\pm(-)^nC\g^0\g^{(n)}\theta_{\mp}~.
\end{equation}
The chirality matrix is defined by
\be
\g_5:=i\g_0\dots\g_3; ~~~~~~\g_5^2=1 ~.
\end{equation}

\subsection{Spinors in $D=10\rightarrow 4+6$}

The charge conjugation matrix in $1+9$ dimensions satisfies
\be
C^{Tr}=-C; ~~~~~~ (C\G^M)^{Tr}=C\G^M~.
\end{equation}
The fundamental (16-dimensional, chiral) spinor representation 
$\e_\pm$ is real and we define the reality condition by
\be
\overline{\e}_\pm=\e_\pm^{Tr}C~.
\end{equation}
%
%
%
The chirality matrix is defined by
\be
\G_{11}:=\G_0\dots\G_9; ~~~~~~\G_{11}^2=1 ~.
\end{equation}
We decompose the ten-dimensional Gamma matrices as
\begin{alignat}{3}
\G^\mu&=\g^\mu\otimes 1, &~~~~~\mu=0,\dots 3&\nn\\
\G^m&=\g_5\otimes \g^m, &~~~~~ m=4\dots 9&~.
\end{alignat}
It follows that
\beal
C_{10}=C_4\g_5\otimes C_6;~~~~~~ \G_{11}=\g_5\otimes\g_7~.
\end{align}

\section{$SU(3)$ structure}
\label{appx}

Using (\ref{fierz}) one can show the following useful 
identities satisfied by $J$, $\Omega$:
\be
\Omega_{[abc}\Omega^*_{def]}=\frac{2i}{5}\varepsilon_{abcdef}
\end{equation}
\be
\Omega_{abc}\Omega^{*def}=48(\Pi^+)_{[a}{}^{[d}
(\Pi^+)_{b}{}^{e}(\Pi^+)_{c]}{}^{f]}
\end{equation}
\be
\Omega_{abc}\Omega^{*ade}=16(\Pi^+)_{[b}{}^{[d}(\Pi^+)_{c]}{}^{e]}
\end{equation}
\be
\Omega_{abc}\Omega^{*abd}=16(\Pi^+)_c{}^d
\end{equation}
\be
|\Omega|^2=48
\end{equation}
\be
\varepsilon_{abcdef}J^{cd}J^{ef}=-8J_{ab}
\end{equation}
\be
\varepsilon_{abcdef}J^{ef}=-6J_{[ab}J_{cd]}
\end{equation}
\be
\varepsilon_{abcdef}=-15J_{[ab}J_{cd}J_{ef]}
\end{equation}
Note that from the last one it follows that 
\be
d Vol_6=-\frac{1}{6}J^3~.
\end{equation}
The following relations are useful 
in analyzing the supersymmetry conditions of section \ref{analysis}.
\beal
0&=(\Pi^+)_m{}^n\gamma_n\eta_-\nn\\
\gamma_{mn}&=iJ_{mn}\eta_++\frac{1}{2}\Omega_{mnp}\gamma^p\eta_-\nn\\
\gamma_{mnp}\eta_-&=-3iJ_{[mn} \gamma_{p]}\eta_--\Omega^*_{mnp}\eta_+~.
\end{align}
The above equations 
together with tensor decompositions (\ref{bexp},\ref{hexp},\ref{gexp}) 
give
\beal
B_{np}
(\g_m{}^{np}-14\delta_m{}^n\g^p)\eta_-
&=\Big(\frac{5i}{3}B^{(0)}\g_{mt}-16\tilde{B}_{mt}-\frac{3}{4}\Omega_{mt}{}^sB_S^{(0,1)}
\Big)\gamma^t\eta_--B_m^{(0,1)}\eta_+\nn\\
H_{npq}
(\g_m{}^{npq}-9\delta_m{}^n\g^{pq})&=\Big(-H^{(0)*}g_{mt}-\frac{9}{2}\Omega_t{}^{pq}\tilde{H}_{mpq}^{(1,2)}
-\frac{3}{2}\Omega_m{}^{pq}\tilde{H}_{tpq}^{(1,2)}
+\frac{3i}{2}\Omega_{mt}{}^pH_p^{(0,1)}
\Big)\gamma^t\eta_-\nn\\
&-(12iH_m^{(1,0)}+6iH_m^{(0,1)}  )\eta_+\nn\\
G_{npqr}(\g_m{}^{npqr}-\frac{20}{3}\delta_m{}^n\gamma^{pqr})
&=\Big(\frac{7}{3}G^{(0)}g_{mt}+32i\tilde{G}_{mt} 
-\frac{1}{3}\Omega_{mt}{}^pG_p^{(0,1)}
\Big)\gamma^t\eta_-
+\frac{20}{3}G_m^{(0,1)}\eta_+
\label{a1}
\end{align}
and
\beal
\gamma^{mn}B_{mn}\eta_-&=-iB^{(0)}\eta_- -\frac{1}{2}B_t^{(0,1)}\gamma^t\eta_+\nn\\
\gamma^{mnp}H_{mnp}\eta_+&=H^{(0)*}\eta_-+3iH^{(0,1)}_t\gamma^t\eta_+\nn\\
\gamma^{mnpq}G_{mnpq}\eta_+&=-3G^{(0)}\eta_--2G^{(0,1)}_t\gamma^t\eta_+
~.
\label{a2}
\end{align}
Finally, one can show the following Hodge-dualizations
\beal
*J&=\frac{1}{2}J\wedge J \wedge dVol_4\nn\\
*(J\wedge J)&=2J\wedge dVol_4\nn\\
*\Omega &= i\Omega\wedge dVol_4\nn\\
*\tilde{B}&=-J\wedge \tilde{B}\wedge dVol_4~,
\label{kokoriko}
\end{align}
where in proving the last one we used the fact that $\tilde{B}$ 
is primitive.

%
%

\end{document}